\documentclass{PoS}

\title{\emph{Swift}/XRT follow-up observations of unidentified \emph{INTEGRAL}/IBIS sources}

\ShortTitle{Swift/XRT follow-up observations of unidentified INTEGRAL/IBIS sources}

\author{\speaker{R. Landi}\\
          INAF/IASF Bologna, Via P. Gobetti 101, 40129 Bologna, Italy\\
          E-mail: \email{landi@iasfbo.inaf.it}}

\author{L. Bassani\\   
        INAF/IASF Bologna, Via P. Gobetti 101, 40129 Bologna, Italy\\
        E-mail: \email{bassani@iasfbo.inaf.it}}

\author{A. Malizia\\
        INAF/IASF Bologna, Via P. Gobetti 101, 40129 Bologna, Italy\\
        E-mail: \email{malizia@iasfbo.inaf.it}}

\author{J. B. Stephen\\
        INAF/IASF Bologna, Via P. Gobetti 101, 40129 Bologna, Italy\\
        E-mail: \email{stephen@iasfbo.inaf.it}}

\author{A. Bazzano\\
        INAF/IASF Roma, Via Fosso del Cavaliere 100, 00133 Roma, Italy\\
        E-mail: \email{angela.bazzano@iasf-roma.inaf.it}}

\author{M. Fiocchi\\
        INAF/IASF Roma, Via Fosso del Cavaliere 100, 00133 Roma, Italy\\
        E-mail: \email{mariateresa.fiocchi@iasf-roma.inaf.it}}

\author{A. J. Bird \\
        School of Physics and Astronomy, University of Southampton, Highfield, SO17 1BJ, UK\\
        E-mail: \email{ajd@phys.soton.ac.uk}}

\abstract{Many sources listed in the 4$^{\rm th}$ IBIS/ISGRI survey are still unidentified, 
i.e. lacking an X-ray counterpart or simply not studied at lower energies ($<$ 10 keV).
The cross-correlation between the list of IBIS sources in the 4$^{\rm th}$ catalogue and the 
\emph{Swift}/XRT data archive is of key importance to search for the X-ray counterparts; in fact, 
the positional accuracy of few arcseconds obtained with XRT allows us to perform more efficient 
and reliable follow-up observations at other wavelengths (optical, UV, radio).
In this work, we present the results of the XRT observations for four new gamma-ray 
sources: IGR J12123--5802, IGR J1248.2--5828, IGR J13107--5626 and IGR J14080--3023.
For IGR J12123--5802 we find a likely counterpart, but further information are needed to classified 
this object, IGR J1248.2--5828 is found to be a Seyfert 1.9, for IGR J13107--5626 we suggest a 
possible AGN nature, while IGR J14080--3023 is classified as a Seyfert 1.5 galaxy.}

\FullConference{The Extreme sky: Sampling the Universe above 10 keV - extremesky2009,\\
		October 13-17, 2009\\
		Otranto (Lecce) Italy}

\begin{document}

\begin{table*}
\begin{center}
\footnotesize
\caption{XRT position and classification of the counterpart of the IBIS sources.}
\label{Tab1}
\begin{tabular}{lccccc}
\hline
\hline
Source & R.A. & Dec & Error & Counterpart & Type \\
    &  (J2000)& (J2000) & (arcsec)  &    &    \\
\hline
\hline
IGR J12123--5802  &  $12^{\rm h}12^{\rm m}25^{\rm s}.97$ & $-58^\circ00^{\prime}23^{\prime \prime}.1$ 
& 3.7 &  2MASS J12122623--5800204 & unidentified  \\
\hline
IGR J1248.2--5828 & $12^{\rm h}47^{\rm m}57^{\rm s}.82$ & $-58^\circ29^{\prime}59^{\prime \prime}.1$ 
& 4.0 & 2MASX J12475784--5829599  & Seyfert 1.9  \\
\hline
IGR J13107--5626  & $13^{\rm h}10^{\rm m}37^{\rm s}.27$ & $-56^\circ26^{\prime}56^{\prime \prime}.7$ 
& 4.4 & 2MASX J13103701--5626551 & AGN candidate  \\
\hline
IGR J14080--3023  & $14^{\rm h}08^{\rm m}06^{\rm s}.57$ & $-30^\circ23^{\prime}52^{\prime \prime}.6$ 
& 3.6 & 2MASX J14080674--3023537  &  Seyfert 1  \\
\hline
\hline
\end{tabular}
\end{center}
\end{table*}

\section{\bf IGR J12123--5802}
\begin{figure}
\centering
\includegraphics[width=0.45\linewidth,angle=0]{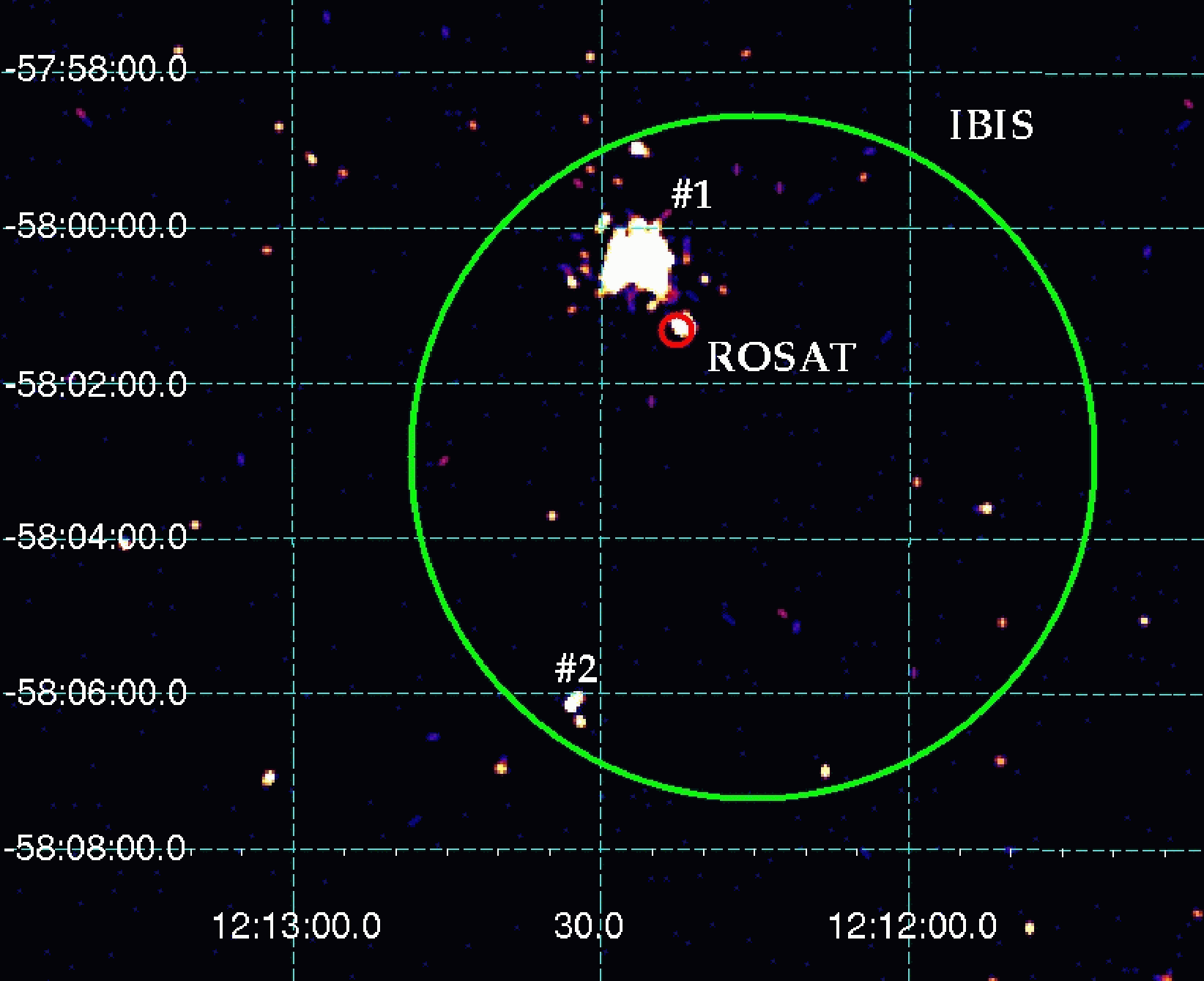}
\hspace{0.5cm}
\includegraphics[width=0.50\linewidth,angle=0]{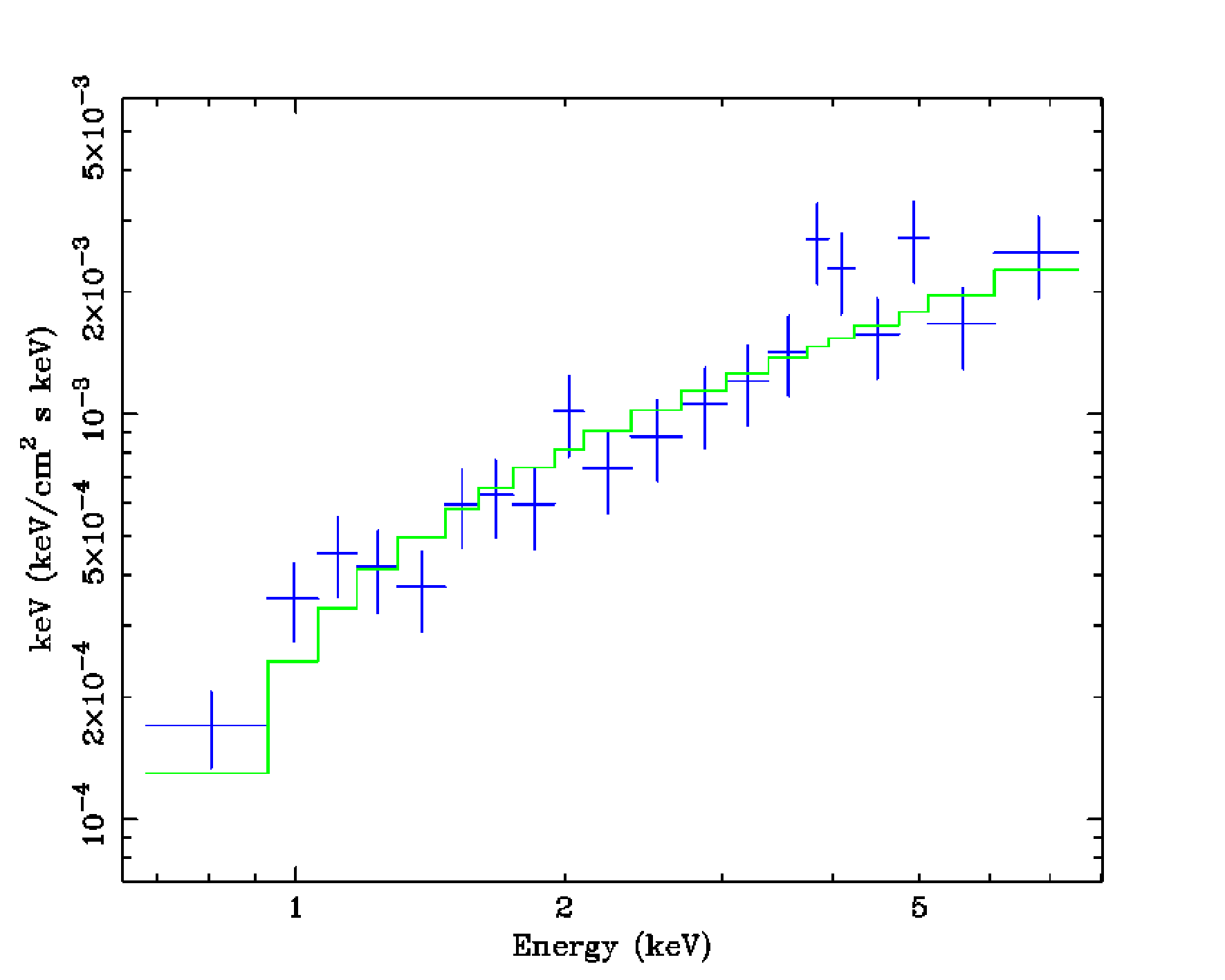}
\caption{\emph{Left panel}: XRT 0.3--10 keV image of the region surrounding IGR J12123--5802.
The larger green circle represents the IBIS position and uncertainty, while the two X-ray sources 
detected within it are labelled as \#1 and \#2. Also plotted is the position (smaller red circle) 
of a ROSAT Bright Survey source (1RXS J121222.7--580118) located $\sim$1$^{\prime}$ away from source 
\#1. \emph{Right panel}: XRT spectrum of source \#1, the likely counterpart of IGR J12123--5802, fitted 
with a power law passing through Galactic absorption.}
\label{fig1}
\end{figure}

Within the IBIS positional uncertainty (see Bird et al. 2010), XRT reveals two X-ray source (see 
Figure~\ref{fig1}, left panel). 
Source \#1 has a counterpart in the United States Naval Observatory (USNO--B1.0, Monet et al. 2003)
catalogue with magnitude $R$ $\sim$14.9--16.5, also 
listed in the 2MASS (2 Micron All Sky Survey, Skrutskie et al. 2006) survey, with magnitudes $J$ 
$\sim$15.4, $H$ $\sim$15.2 and $K$ $\sim$15.1. 
For source \#2 (located at R.A.(J2000) = $12^{\rm h}12^{\rm m}32^{\rm s}.40$, Dec(J2000) =
$-58^\circ06^{\prime}09^{\prime \prime}.5$, error radius 6$^{\prime\prime}$.0 and only detected below 4 
keV) we did not find a counterpart in any database, thus suggesting that deeper X-ray observations are 
needed to investigate its nature. 

As can be seen in the left panel of Figure~\ref{fig1}, at $\sim$1$^{\prime}$ away from source \#1 
there is also the ROSAT 
source 1RXS J121222.7--580118, which is detected at $\sim$3$\sigma$ but only at soft energies ($<$ 
3 keV). Although this source belongs to the ROSAT Bright survey, during the XRT pointing 
appears weak, having a 2--10 keV flux of $\sim$$3 \times10^{-13}$ erg cm$^{-2}$ s$^{-1}$ assuming a 
power law model with the photon index frozen to 1.8. This behaviour seems to indicate a variable 
nature for this source, thus making unlikely its association with the IBIS detections, which  
instead is listed as persistent in the 4$^{\rm th}$ survey. From the above considerations, we conclude 
that source \#1 is the likely candidate for the IBIS detection.

The X-ray spectrum of source \#1 (see Figure~\ref{fig1}, right panel) is well modelled with a 
power law passing through Galactic absorption (Kalberla et al. 2005) and having a flat photon index 
($\Gamma$ $\sim$1.3) and a 2--10 keV flux of $\sim$$4.5\times10^{-12}$ erg cm$^{-2}$ s$^{-1}$ (see 
Table~\ref{Tab2}). Although we were able to find the X-ray and optical counterpart of IGR 
J12123--5802, the information collected so far do not allow us to draw any conclusion about its class.
Therefore, optical follow-up observations are needed to determine the real nature of this 
new gamma-ray source.

\begin{figure}
\centering
\includegraphics[width=0.45\linewidth,angle=0]{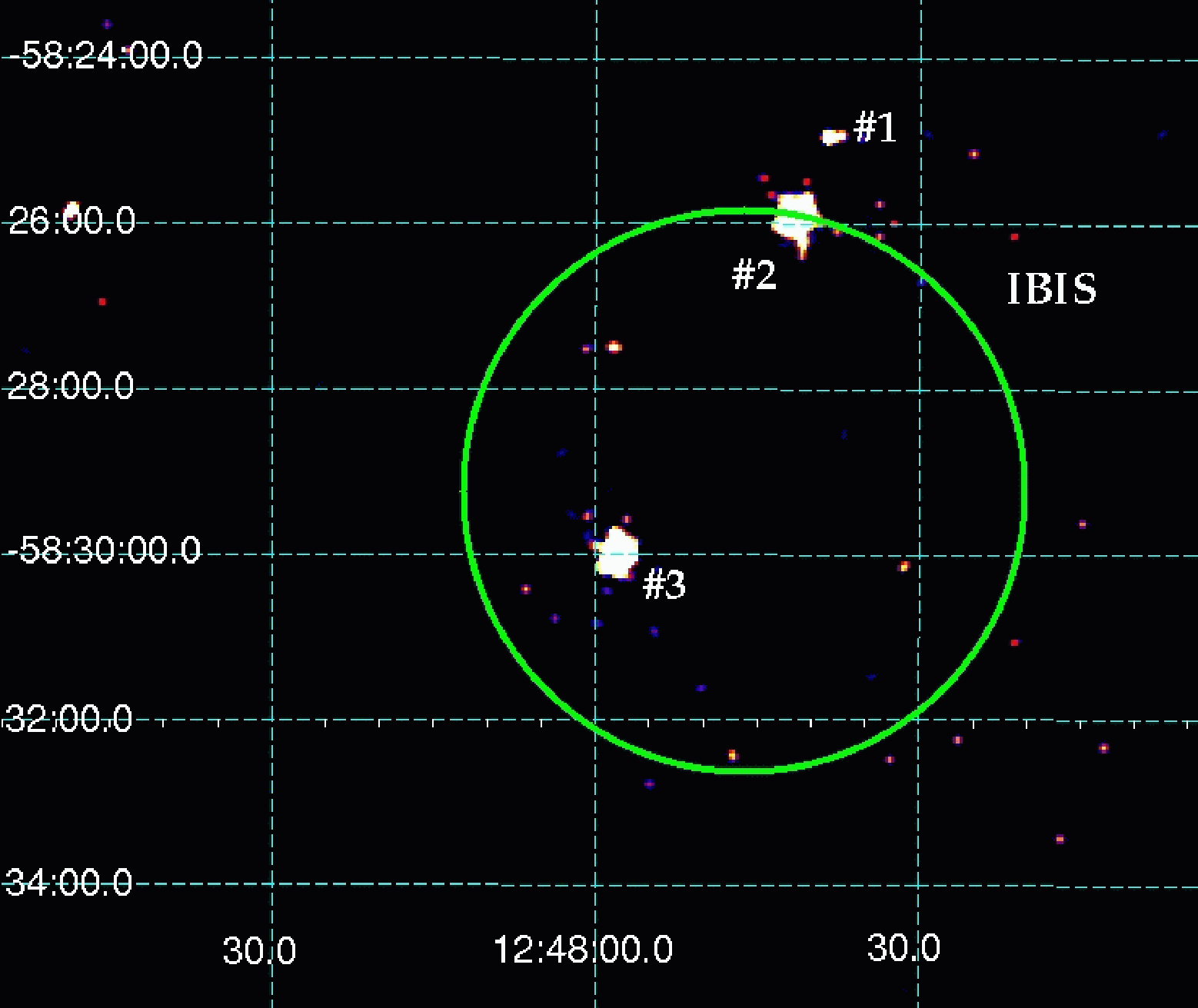}
\hspace{0.5cm}
\includegraphics[width=0.50\linewidth,angle=0]{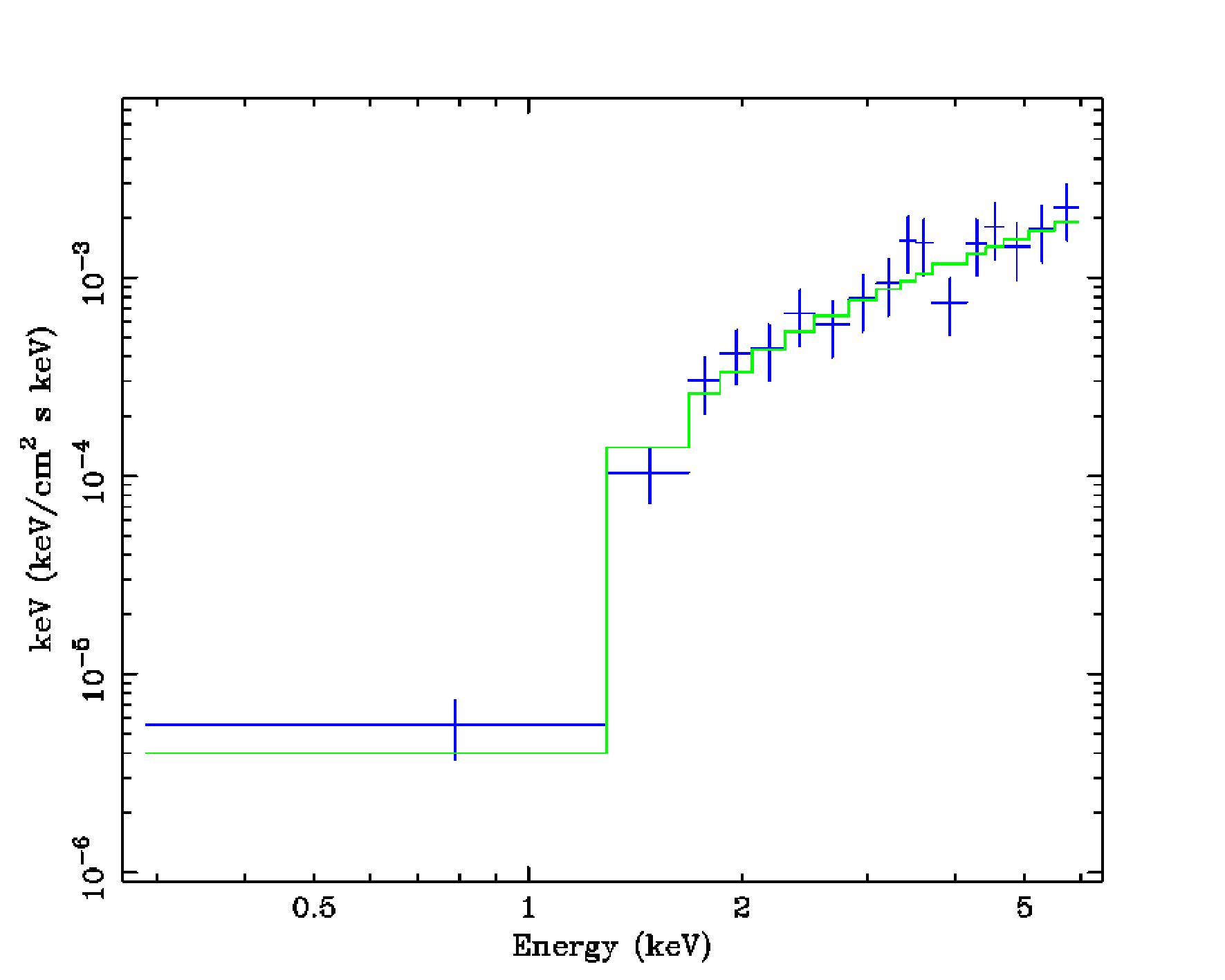}
\caption{\emph{Left panel}: XRT 0.3--10 keV image of the region surrounding IGR J1248.2--5828.
The green circle represents the 90\% IBIS error circle, while the two X-ray sources detected by XRT 
within and at the border of it are labelled as \#2 and \#3. Source \#1 is the fainter
source detected by XRT within the 99\% IBIS error circle. \emph{Right panel}: XRT spectrum of source \#3, 
the likely counterpart of IGR J1248.2--5828, fitted with an absorbed power law.}
\label{fig2}
\end{figure}

\section{\bf IGR J1248.2--5828}

In this case, three X-ray sources are detected within the IBIS positional uncertainty (Bird et al. 2010) 
as can 
be seen in Figure~\ref{fig2}, left panel):
source \#1 is detected within the 99\% IBIS error circle, while source \#2 and \#3 are located at the 
border and inside of the 90\% IBIS uncertainty, respectively. 
Source \#1, located at R.A.(J2000) = $12^{\rm h}47^{\rm m}38^{\rm s}.10$,  Dec(J2000) = 
$-58^\circ24^{\prime}54^{\prime \prime}.3$, error radius = 6$^{\prime \prime}$.0, is the faintest 
object, being detected only at $\sim$3.3$\sigma$ in 0.3--10 keV, and its position is consistent with
an object classified as a star (HIP 62427).
The XRT position of source \#2 (R.A.(J2000) = $12^{\rm h}47^{\rm m}41^{\rm s}.57$, Dec(J2000) = 
$-58^\circ25^{\prime}53^{\prime\prime}.6$, error radius 3$^{\prime \prime}$.9), which is the brightest 
object at soft energies, coincides with a double or multiple star CCDM J12477--5826AB and with a ROSAT 
Faint Survey source (1RXS J124742.1--582544), not yet classified.
The only object still visible above 3 keV is source \#3; it has a counterpart in a 
USNO--B1.0 source having magnitude $R=14.6-15.1$, and it is also listed in the 2MASS Extended survey 
(2MASX J12475784--5829599). The X-ray position is also compatible with a radio source 
belonging to the MGPS--2 (Molonglo Galactic Plane Survey 2nd Epoch Compact Source, Murphy et al. 
2007) catalogue, with a 36 cm flux of $16.5\pm1.1$ mJy.

Source \#1 and \#2 have very soft X-ray spectra well modelled with a thermal bremsstrahlung with 
$kT$ $\sim$0.4 keV and a 2--10 keV flux of $\sim$$2.1 
\times10^{-15}$ erg cm$^{-2}$ s$^{-1}$ and $\sim$$2.5 \times10^{-14}$ erg cm$^{-2}$ s$^{-1}$,
respectively. These characteristics combined with their coincidence with stellar objects makes 
unlikely their associations with the IBIS source.
The X-ray spectrum of source \#3 (see Figure~\ref{fig2}, right panel) is fitted with a flat ($\Gamma$ 
$\sim$0.9) slightly absorbed power law, having a 2--10 keV 
flux of $\sim$$4\times 10^{-12}$ erg cm$^{-2}$ s$^{-1}$ (see 
Table~\ref{Tab2}). The fit is still acceptable if we fix the photon index to 1.8 and provides a high 
intrinsic absorption of $\sim$$2\times 10^{22}$ cm$^{-2}$. This source shows a flux variability of a 
factor of 2.5 on a time-scale of a few months during XRT pointings.
These properties suggest its associations with IGR J1248.2--5828 and, in particular, its 
characteristics in radio and infrared seem to indicate that we are dealing with an AGN. 
This suggestion has recently been confirmed by optical follow-up observations (Masetti et al. in preparation).

\begin{table*}
\begin{center}
\footnotesize
\caption{\emph{Swift}/XRT spectral analysis results of the averaged spectra. Frozen parameters are 
written in square brackets; errors are given at the 90\% confidence level.}
\label{Tab2}
\begin{tabular}{lccccc}
\hline
\hline
Source & $N_{\rm H(Gal)}^{a}$ & $N_{\rm H}^{a}$ & $\Gamma$ & $\chi^2/\nu$ & $F_{\rm (2-10~keV)}^{b}$ \\
\hline
\hline
IGR J12123--5802 & 0.325 & -- & $1.26^{+0.18}_{-0.16}$ & 19.7/18  & $0.45\pm0.02$ \\
\hline
IGR J1248.2--5828 (\#3)    & 0.297    & $0.92^{+0.67}_{-0.82}$ &  $0.86^{+0.76}_{-0.70}$ & 
9.4/14 & $0.41\pm0.03$ \\ 
\hline
IGR J13107--5626 (\#1)     & 0.244    & $39.3^{+24.4}_{-13.3}$ & [1.8] & 7.5/9 & $0.11\pm0.02$ \\
\hline
IGR J14080--3023$^{c}$ (\#1) & 0.0362 & -- & $1.41\pm0.06$ & 144.0/134 & $0.64\pm0.01$  \\
\hline
\hline
\end{tabular}
\end{center}
$^{a}$ In units of $10^{22}$ cm$^{-2}$ (from Kalberla et al. (2005));\\
$^{b}$ In units of $10^{-11}$ erg cm$^{-2}$;\\
$^{c}$ Best-fit model includes a blackbody component with a $kT = 81^{+6}_{-5}$ eV
to account for the excess observed below 2 keV.
\end{table*}

\begin{figure}
\centering
\includegraphics[width=0.45\linewidth,angle=0]{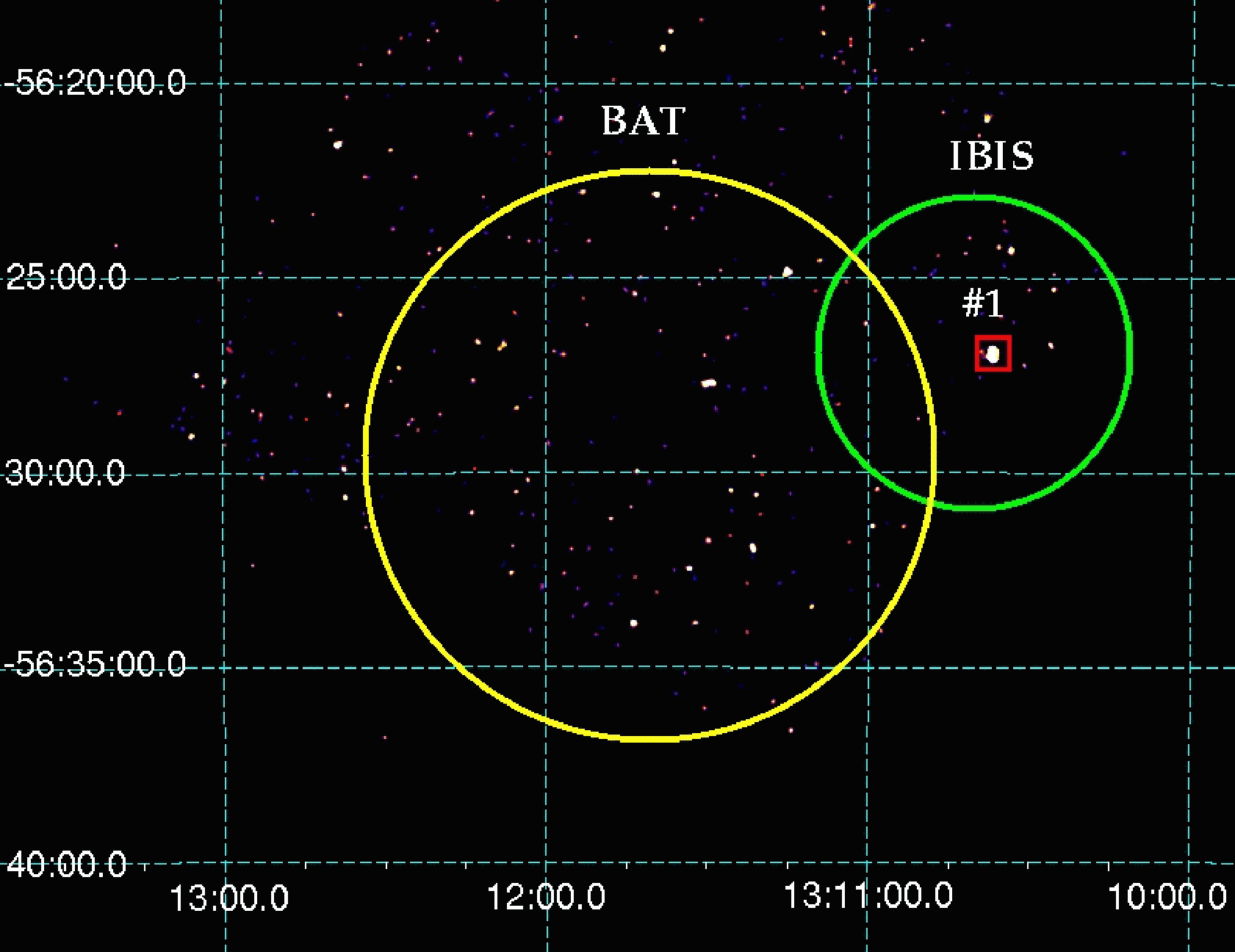}
\hspace{0.5cm}
\includegraphics[width=0.50\linewidth,angle=0]{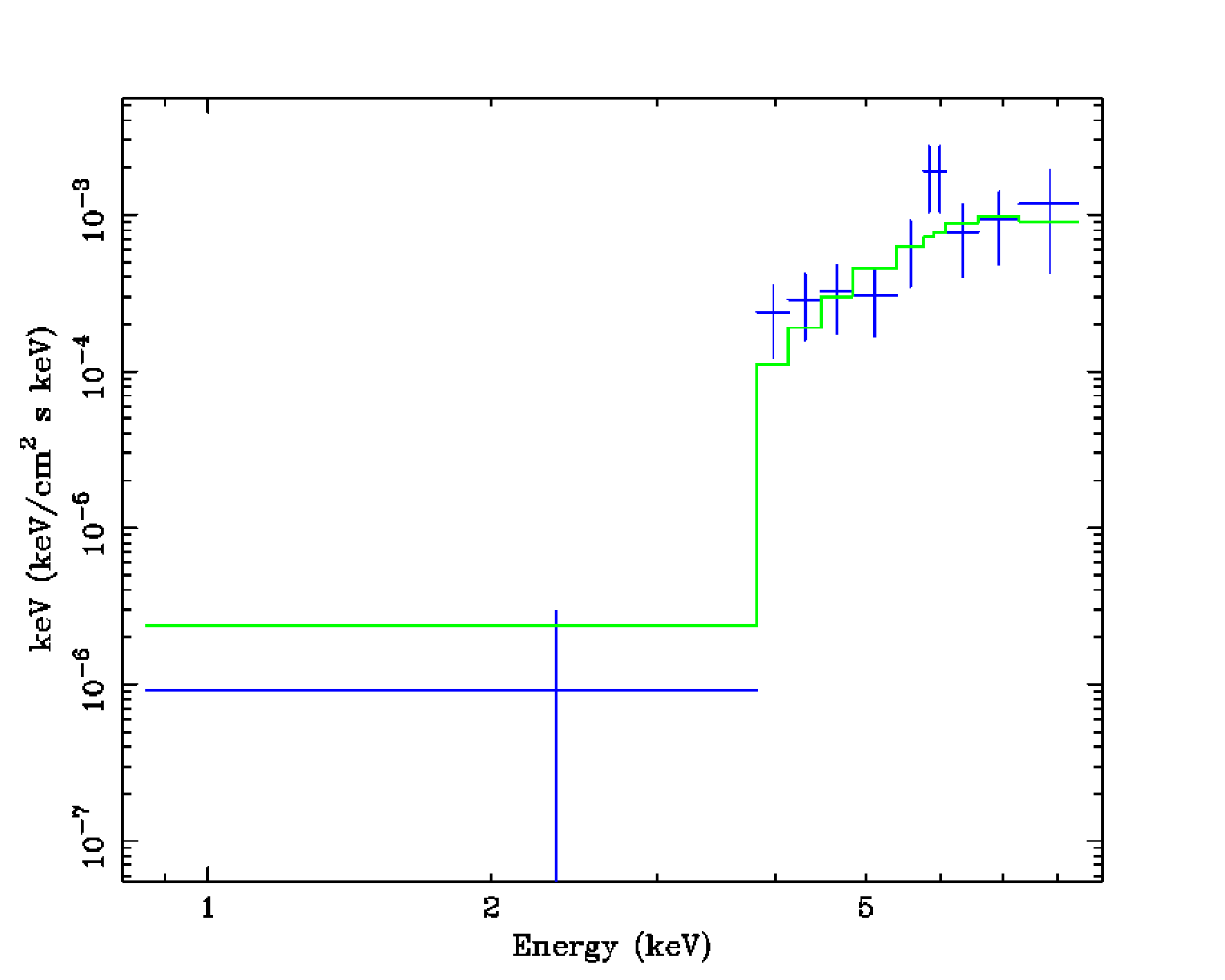}
\caption{\emph{Left panel}: XRT 0.3--10 keV image of the region surrounding IGR J13107--5626.
The green and yellow circles represent the IBIS and BAT position and 
uncertainty, respectively, while the red box shows the location of the 2MASS Extended object (2MASX 
J13103701--5626551) proposed as counterpart. \emph{Right panel}: XRT spectrum of source \#1, the proposed
counterpart of IGR J13107--5626, fitted with an absorbed power law.}
\label{fig3}
\end{figure}

\section{\bf IGR J13107--5626}

There is only one source detected in the whole XRT field of view, which is well located in 
the middle of the IBIS error circle (Bird et al. 2010). It has a counterpart in a
2MASS Extended object (2MASX J13103701--5626551) pointing to a galaxy still unclassified.
It is also listed in the USNO--B1.0 catalogue with magnitude $R=16.6-17.2$.
The XRT position is compatible with a radio source belonging to
the MGPS--2 catalogue with a 36 cm flux of $35.4\pm1.6$ Jy.
As shown in Figure~\ref{fig3} (left panel), despite the partial overlap between the IBIS uncertainty 
and the BAT error circle of Swift J1312.1--5631, the 2MASS Extended object, proposed 
by Tueller et al. (2010) as the counterpart of the BAT source, is well located within the IBIS 
uncertainty, but it is $\sim$$1^{\prime}.9$ away from the border of the BAT error circle and 
$\sim$$9^{\prime}.4$ from the BAT centroid position. 
This makes less convincing the association between the BAT detection and the 
2MASS Extended galaxy, while is more clear its connection with the IBIS detection. 

If we fit the XRT spectrum (see Figure~\ref{fig3}, right panel) with an absorbed power law by fixing 
the photon index to 1.8, we find an
absorption in excess to the Galactic one of $\sim$$4\times 10^{23}$ cm$^{-2}$ and a 2--10 keV flux 
of $\sim$$1\times 10^{-12}$ erg cm$^{-2}$ s$^{-1}$ (see Table~\ref{Tab2}).

Based on the multiwaveband characteristics, we propose for IGR J13107--5626 an absorbed AGN 
classification.

\begin{figure}
\centering
\includegraphics[width=0.45\linewidth,angle=0]{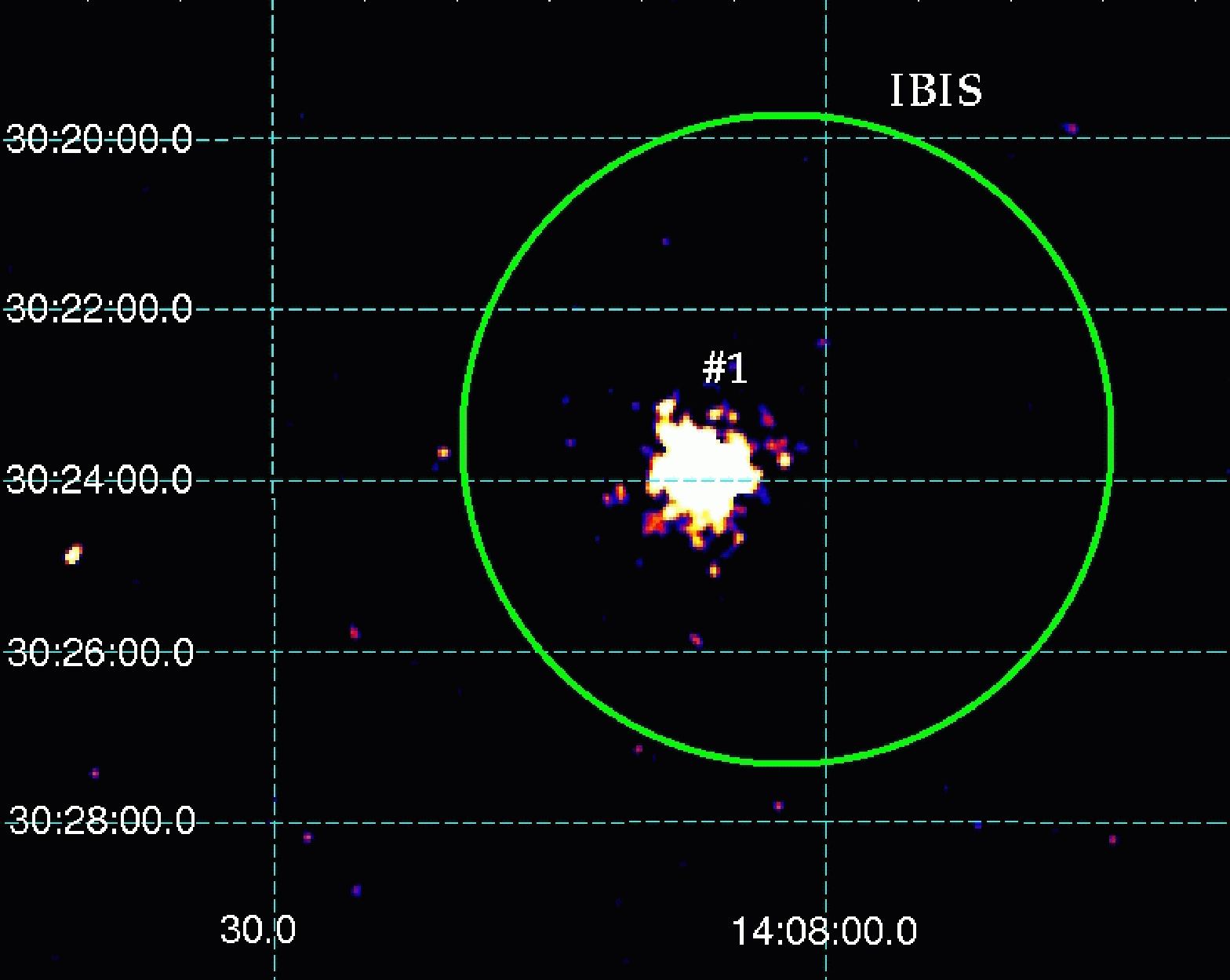}
\hspace{0.5cm}
\includegraphics[width=0.50\linewidth,angle=0]{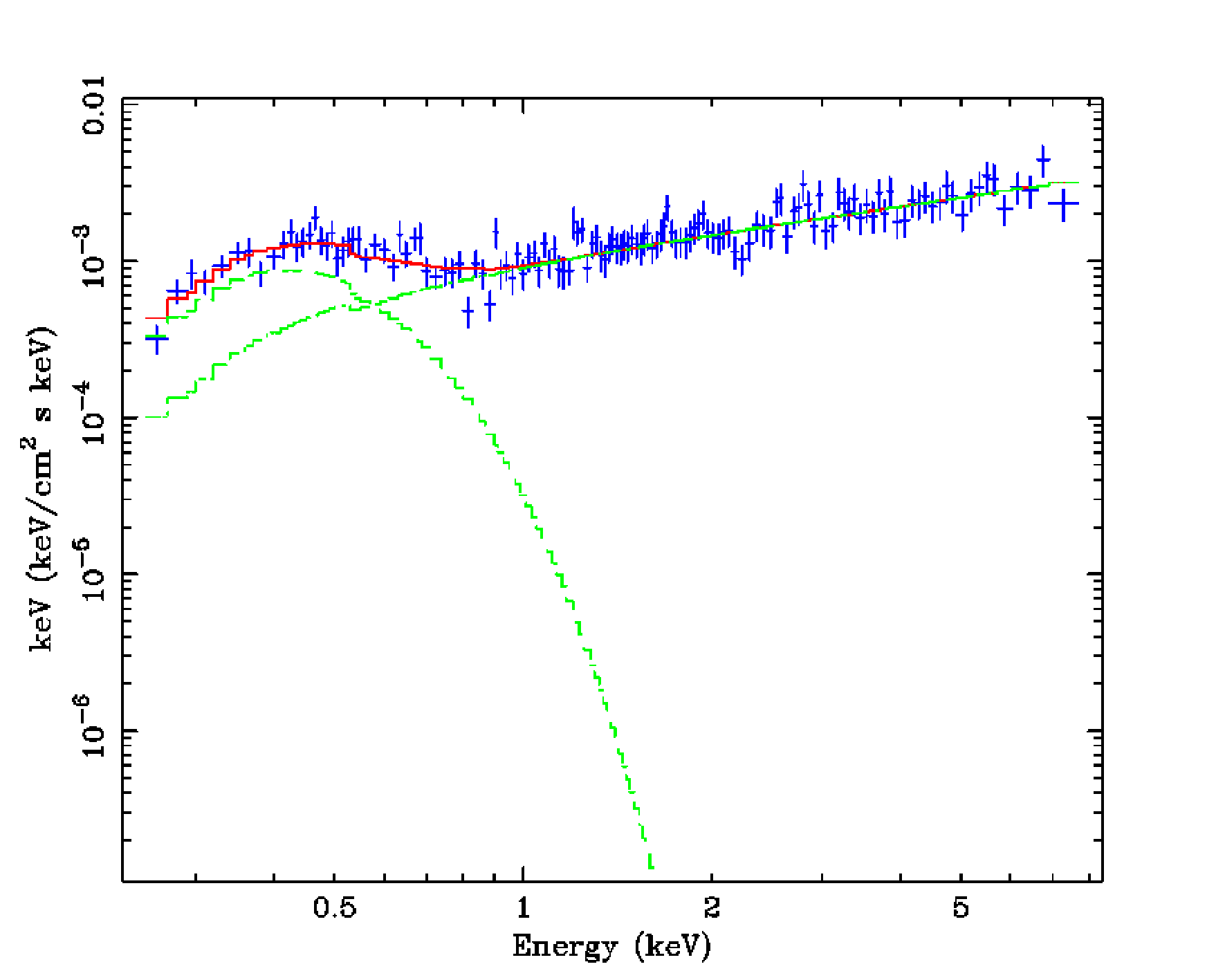}
\caption{\emph{Left panel}: XRT 0.3--10 keV image of the region surrounding IGR J14080--3023. Only 
one X-ray object (source \#1) is detected by XRT within the IBIS uncertainty. \emph{Right panel}: 
XRT spectrum of source \#1, the proposed counterpart of IGR J14080--3023, fitted with a simple power 
law passing through Galactic absorption plus a blackbody component to account 
for the soft X-ray emission observed below 2 keV.} 
\label{fig4}
\end{figure}

\section{\bf IGR J14080--3023}

This source recently appeared in the \emph{Swift}/BAT survey of Cusumano et al. (2010) where it is
associated with a 2MASS Extended object (2MASX J14080674--3023537) classified as a Seyfert 1.5 
(V\'eron-Cetty \& V\'eron 2001).
Indeed, the only source detected by XRT within the IBIS error circle (Bird et al. 2010), as shown in 
Figure~\ref{fig4}, left panel, coincides with the
infrared object, which has a counterpart in the USNO--B1.0 catalogue with magnitude $R=13.7-14.2$ and 
it is also listed in the \emph{XMM-Newton} Slew Survey (XMMSL1 J140806.7--302348)
with a 0.2--12 keV flux of $3.2 \times 10^{-12}$ erg cm$^{-2}$ s$^{-1}$, lower than the range measured 
during the XRT pointings ($\sim$($0.9-1.3) \times 10^{-11}$ erg cm$^{-2}$ s$^{-1}$) in the same energy 
band.

The fit of the X-ray spectrum, in addition to the power law passing through Galactic absorption 
(Kalberla et al. 2005), requires a blackbody component
to account for the presence of soft emission below 2 keV (see Figure~\ref{fig4}, right panel and 
Table~\ref{Tab2}).  
This best-fit model provides a photon index $\Gamma$ $\sim$1.4 and a blackbody temperature $kT$ 
$\sim$81 eV. No extra absorption in excess to the Galactic one is required by the data.
The source shows a flux variability of a factor of 1.3 with a time-scale of a few months, but no 
changes in the spectral shape are observed.
The X-ray properties are compatible with the AGN classification proposed for this source.

\section{\bf Conclusions}
In this work, we show how follow-up observations in X-rays are of key importance to search for 
counterparts of high energy emitters. The cross-correlation between the 4$^{\rm th}$ IBIS 
catalogue and the \emph{Swift}/XRT data archive allowed us to pinpoint unambiguously the counterpart 
of four IBIS sources. IGR J1248.2--5828 is classified as a Seyfert 1.9 galaxy, while the X-ray 
properties of IGR J14080--3023 are compatible with its classification as a Seyfert 1.5 galaxy; based 
on its X-ray characteristics, IGR J13107--5626 is likely an AGN, while for IGR J12123--5802 the data 
available so far do not allow us to assess its nature and only optical measurements are needed to 
firmly establish the nature of this new gamma-ray source.

\end{document}